%Paper: hep-th/9211137
%From: A.Kirillov@newton.cam.ac.uk
%Date: Mon, 30 Nov 92 14:32 GMT

\font\tenbf=cmbx10

\font\eightrm=cmr8
\font\eightit=cmti8

\def\sectiontitle#1\par{\vskip0pt plus.1\vsize\penalty-250
 \vskip0pt plus-.1\vsize\bigskip\vskip\parskip
 \message{#1}\leftline{\tenbf#1}\nobreak\vglue 5pt}
\def\qed{\hbox{${\vcenter{\vbox{
    \hrule height 0.4pt\hbox{\vrule width 0.4pt height 6pt
    \kern5pt\vrule width 0.4pt}\hrule height 0.4pt}}}$}}
\hsize=5.0truein
\vsize=7.8truein
\parindent=15pt
\nopagenumbers
\baselineskip=10pt
\vglue 5pc
\baselineskip=13pt
\headline{\ifnum\pageno=1\hfil\else%
{\ifodd\pageno\rightheadline \else \leftheadline\fi}\fi}
\def\rightheadline{\hfil\eightit
Dilogarithm identities and  spectra in conformal field theory
\quad\eightrm\folio}
\def\leftheadline{\eightrm\folio\quad
\eightit
Anatol N. Kirillov
%! Running author (even)
\hfil}
\voffset=2\baselineskip
\centerline{\tenbf
DILOGARITHM \hskip 0.1cm IDENTITIES \hskip 0.1cm AND
 }
\centerline{\tenbf
SPECTRA \hskip 0.1cm IN \hskip 0.1cm CONFORMAL \hskip 0.1 cm
FIELD \hskip 0.1cm THEORY }
\vglue 24pt
\centerline{\eightrm
ANATOL N. KIRILLOV
%! AUTHOR:
}
\baselineskip=12pt
\centerline{\eightit
Isaac Newton Institute for Mathematical Sciences,
%! Use the permanent address (University Name, etc.)
}
\baselineskip=10pt
\centerline{\eightit
20 Clarkson Road, Cambridge, CB3 OEH, U.K.
}
\baselineskip=12pt
\centerline{\eightit
and }
\baselineskip=12pt
\centerline{\eightit
Steklov Mathematical Institute,
%! Use the permanent address (University Name, etc.)
}
\baselineskip=10pt
\centerline{\eightit
Fontanka 27, St.Petersburg, 191011, Russia
%! The permanent address (City, State ZIP/Zone, Country)
}
%More than two authors, see the sample prints.
\vglue 20pt
\centerline{  }
\centerline{\eightrm ABSTRACT}
{\rightskip=1.5pc
\leftskip=1.5pc
\eightrm\parindent=1pc
We prove new identities between the values of Rogers dilogarithm function and
describe a connection between these identities and spectra in conformal field
theory.
\vglue12pt}
\baselineskip=13pt
\overfullrule=0pt
%-------------------------------------------------------------------

\def\qed{\hfill$\vrule height 2.5mm width 2.5mm depth 0mm$}
\vskip 1.5cm

The dilogarithm function defined for $0\le z\le 1$ by
$$Li_2(z)=\sum_{n=1}^{\infty}{z^n\over n^2}=-\int_0^z{\log (1-t)\over t}dt,$$
is one of the lesser transcendental functions. Nonetheless, it has many
intriguing properties and has appeared in various branches of mathematics and
physics. Here we mention only the appearences of dilogarithm function in
number theory (the study of asymptotic behaviour of partitious [RS]; the
values of $\zeta$-functions in some special points [Za]) in algebraic
$K$-theory (A.Beilinson, S.Bloch, A.Groncharov,...).

In physics, the dilogarithm appears at first from a calculation of magnetic
susceptibility in the $XXZ$ model at small magnetic field [KR]. More recently
[Z], the dilogarithm identities (through the Thermodynamic Bethe Ansatz (TBA))
through appear in the context of investigation of $UV$ limit or the
critical behaviour of integrable 2-dimensional quantum field theories and
lattice models [Z], [DR], [KP],...). One aim of this paper is to prove some
new identities between the values of Rogers dilogarithm function and to show
that any rational number may be obtained as the value of some dilogarithm
sum. More detailes and further results will appear in [Ki2].
\medbreak
%\vskip 0.5cm
\vfill\eject

{\bf \S 1.~Definition~and~the~basic~properties~of~Rogers~dilogarithm.}
\medbreak

Let us remind the definition of the Rogers dilogarithm function $L(x)$ for
\break
$x\in (0,1)$
$$
L(x)=-{1\over 2}\int_0^x\left[ {\log (1-x)\over x}+{\log x\over 1-x}\right]dx=
\sum_{n=1}^\infty {x^n\over n^2}+{1\over 2}\log x\cdot \log(1-x).
%~~0\le x\le 1.
\eqno (1.1)
$$

The following two classical results (see e.g. [Le], [GM], [Ki]) contain the
basic properties of the function $L(x)$.
\medbreak

{\bf Theorem~A}.~~The function $L(x)\in C^{\infty}((0,1))$ and satisfies the
following functional equations
$$\eqalignno{\hfil &1.~~L(x)+L(1-x)={\pi^2\over 6},~~0<x<1,
%L(1),~~
%x\in\bf R
& (1.2)\cr
&2.~~L(x)+L(y)=L(xy)+L\left({x(1-y)\over 1-xy}\right)+
L\left({y(1-x)\over 1-xy}\right),& (1.3)}
$$
where $0<x,~y<1$.
\medbreak

{\bf Theorem~B}.~~Let$f(x)$ be a function of class $C^3((0,1))$ and satisfies
the  relations (1.2) and (1.3). Then we have
$$ f(x)={\rm const}\cdot L(x)$$

We continue the function $L(x)$ on all real axis
${\bf R}={\bf R}^1\cup\{\pm\infty\}$ by the following rules
$$\eqalignno{
& L(x)={\pi^2\over 3}-L(x^{-1}),~~{\rm if}~~x>1, &(1.4)\cr
& L(x)=L\left({1\over 1-x}\right) -{\pi^2\over 6},~~{\rm if}~~x<0,& (1.5)\cr
& L(0)=0,~~L(1)={\pi^2\over 6},~~L(+\infty )={\pi^2\over 3},~~L(-\infty )=
-{\pi^2\over 6}.}
$$

The present work will concern with relations between the values of the
Rogers dilogarithm function at certain algebraic numbers. More exactly, let us
consider an abelian subgroup $W$ in the field of rational numbers ${\bf Q}$
$$
W=\left \{ \sum_i{n_iL(\alpha_i)\over L(1)}~|~n_i\in{\bf Z},~~
\alpha_i\in{\overline {\bf Q}}\cap {\bf R}~~{\rm for~all}~~
i\right \} \cap{\bf Q}.
\eqno (1.6)
$$

According to a conjecture of W. Nahm [Na] the abelian group $W$ ``coincides''
with the spectra in rational conformal field theory.
Thus it seems very interesting
task  to obtain more explicit description of the group $W$ (e.g. to find a
system of generators for $W$) and also to connect already known results about
 the spectra in conformal field theory (see e.g. [BPZ], [GKO], [FF], [FQS],
[Ka], [KW], [KP]) with suitable
elements in $W$. One of our main results of the present paper allows to
describe some part of a system of generators for abelian group $W$. As a
corollary, we will show that the spectra of unitary minimal models
[BPZ] and some
others are really contained in $W$.
%\vskip 0.5cm
\vfill\eject
{\bf \S 2.~~Basic~identities~and~conformal~weights.}
\medbreak

In this section we present our main results dealing with a computation of the
following dilogarithm sum
$$
\sum_{k=1}^{n-1}\sum_{m=1}^rL\left({\sin k\varphi\cdot
\sin (n-k)\varphi\over\sin
 (m+k)\varphi\cdot\sin (m+n-k)\varphi}\right)
%={(n^2-1)r\over r+n}\cdot {\pi^2\over 6}
:={\pi^2\over 6}s(j,n,r), \eqno (2.1)
$$
 where $ \varphi =\displaystyle{{(j+1)\pi\over n+r}},~~0\le j\le n+r-2$.

It is clear that $s(j,n,r)=s(n+r-2-j,n,r)$, so we will assume in sequel that
$0\le 2j\le n+r-2$.

The dilogarithm sum (2.1) corresponds to the Lie algebra of type $A_{n-1}$.
The case $j=0$ was considered in our previous paper [Ki], where it was proved
that $s(0,n,r)=\displaystyle{{(n^2-1)r\over n+r}}$.
It was stated in [Ki] that this number
coincides with the central charge of the $SU(n)$ level $r$ WZNW model.
Before formulating our result about computation of the sum $s(j,n,r)$ let us
remind the definition of Bernoulli polynomials. They are defined by the
generating function
$${te^{xt}\over e^t-1}=\sum_{n=0}^{\infty}B_n(x){t^n\over n!},~~|t|<2\pi.
$$
We use also modified Bernoulli polynomials
$$ {\overline B}_n(x)=B_n(\{ x\} ),~~{\rm where}~~\{ x\} =x-[x]
$$
be a fractional part of $x\in {\bf R}$. It is well-known that
$$\eqalignno{
&{\overline B}_{2n}(x)=(-1)^n{2(2n)!\over (2\pi )^{2n}}\sum_{k=1}^{\infty }
{\cos 2k\pi x\over k^{2n}},& (2.2)\cr
&{\overline B}_{2n+1}(x)=(-1)^{n-1}{2(2n+1)!\over (2\pi )^{2n+1}}
\sum_{k=1}^{\infty }{\sin 2k\pi x\over k^{2n+1}}.}
$$
%\medbreak

{\bf Theorem~2.1.}~~We have
$$s(j,n,r)=6(r+n) \sum_{k=0}^{[{n-1\over 2}]}\left\{
{1\over 6}-{\overline B}_2\biggl( (n-1-2k)
\theta\biggr) \right\} -{1\over 4}\biggl\{ 2n^2+1+3(-1)^n\biggr\},\eqno (2.3)
$$
where $\theta =\displaystyle{{j+1\over r+n}}$ and ${\rm g.c.d.}(j+1,r+n)=1$.
%\medbreak

{\bf Theorem~2.2.}~(level-rank duality)
$$s(j,n,r)+s(j,r,n)=nr-1. \eqno (2.4)
$$
%\medbreak

{\bf Corollary~2.3.}~~We have
$$s(j,n,r)=c_r^{(n)}-24h_j^{(r,n)}+6\cdot{\bf Z}_+, \eqno (2.5)
$$
where
$$c_r^{(n)}={(n^2-1)r\over n+r},~~h_j^{(r,n)}=
{n(n^2-1)\over 24}\cdot{j(j+2)\over r+n},~~0\le j\le r+n-2,\eqno (2.6)
$$
are the central charge and conformal dimensions of the $SU(n)$ level $r$ WZNW
primary fields, respectively.

Proof.~~Let us remind that
$$B_2(x)=x^2-x+{1\over 6}~~{\rm and}~~{\overline B}_2(x)=B_2(\{ x\} ). $$
Thus,
$$\eqalignno{
&6(r+n)\sum_{k=0}^{[{n-1\over 2}]}\left({1\over 6}-{\overline B}_2\biggl(
(n-1-2k)\theta\biggr) \right) =\cr
&=6(r+n)\sum_{k=0}^{[{n-1\over 2}]}(n-1-2k)\theta -6(r+n)\sum_{k=0}^{[{n-1
\over 2}]}(n-1-2k)^2\theta ^2+\cr
&+6(r+n)\sum_{k=0}^{[{n-1\over 2}]}
\biggr[(n-1-2k)\theta\biggl]\biggl((n-1-2k)\theta-1+\biggl\{ (n-1-2k)\theta
\biggr\}\biggr):=\cr
&~~\cr
& =6\Sigma_1-6\Sigma_2+6\Sigma_3. }
$$
Now if we take $\theta =\displaystyle{{j+1\over r+n}}$, then  it is clear that
$\Sigma_3\in{\bf Z}_+$. In order to compute $\Sigma_1$ and $\Sigma_2$ we use
the following summation formulae
$$\eqalignno{
&\sum_{k=0}^{[{n-1\over 2}]}(n-1-2k)={2n^2-1+(-1)^n\over 8}=
\left[{n\over 2}\right]\cdot\left[ {n+1\over 2}\right],\cr
&\sum_{k=0}^{[{n-1\over 2}]}(n-1-2k)^2={n(n^2-1)\over 6}.}
$$
Consequently ~~ $s(j,n,r)=$
$$\eqalignno{
&={3(2n^2-1+(-1)^n)(j+1)\over 4}-{n(n^2-1)(j+1)^2\over r+n}-{2n^2+1+
3(-1)^n\over 4}+6\Sigma_3=\cr
&={(n^2-1)r\over r+n}-{n(n^2-1)j(j+2)\over r+n}+6j
\left[{n\over 2}\right]\cdot\left[ {n+1\over 2}\right] +6\Sigma_3.}
$$
\qed

For small values of $j$ we may compute the sum in (2.3) and thus to find
corresponding positive integer in (2.5).
%\medbreak

{\bf Corollary~2.4.}

$i)$ if ~$j\le r$,~ then
$$s(j,2,r)=c_r^{(2)}-24h_j^{(r,2)}+6j={3r\over r+2}+6~{j(r-j)\over r+2}.
\eqno (2.7)
$$

$ii)$ if ~$2j\le r+1$, ~ then
$$s(j,3,r)={8r\over r+3}-24~{j(j+2)\over r+3}+12j. \eqno (2.8)
$$

$iii)$ if ~$(n-1)j<r+1$,~ then
$$s(j,n,r)=c_r^{(n)}-24h_j^{(r,n)}+6j\cdot \left[{n\over 2}\right]
\cdot\left[{n+1\over 2}\right]. \eqno (2.9)
$$

Proof.~~An assumption ~$(n-1)j<r+1$~ is equivalent to a condition\break
$\displaystyle{{(n-1)(j+1)\over n+r}}<1$. So the term $\Sigma_3$
(see a proof of Corollary 2.3) is equal to zero.
\qed

It seems interesting to find a meaning of the positive integer in (2.5).
Now we want to find a ``dilogarithm interpretation'' of the central charges
and conformal dimensions for some well-known conformal models.
\medbreak

{\bf Corollary~2.5.}~~We have
$$s(j_1,2,k)+s(0,2,1)-s(j_2,2,k+1)=c_k-24h_{j_1+1,j_2+1}+6(j_1-j_2)(j_1-j_2+1),
\eqno (2.10)
$$
where
$$c_k=1-{6\over (k+2)(k+3)},~~h_{r,s}^{(k)}={[(k+3)r-(k+2)s]^2-1
\over 4(k+2)(k+3)} \eqno (2.11)
$$
are the central charge and conformal dimensions of the primary fields for
unitary minimal conformal models [BPZ], [Ka], [FQS].
%\medbreak

{\bf Corollary~2.6.}
$$s(j_1,n,k)+s(0,n,1)-s(j_2,n,k+1)=c_{k,n}-24h_{j_1+1,j_2+1}^{(n)}+12{\bf Z}_+,
\eqno (2.12)
$$
where
$$\eqalignno{
&c_{k,n}=(n-1)\left\{ 1-{n(n+1)\over (k+n)(k+n+1)}\right\} ,& (2.13)\cr
&h_{r,s}^{(n)}(k)={n(n^2-1)\over 24}\cdot{[(k+n+1)r-(k+n)s]^2-1\over
(k+n)(k+n+1)}}
$$
are the central charge and conformal dimensions of the primary fields for $W_n$
models [Bi].

Finally we give a ``dilogarithm interpretation'' for the central charge and
conformal weights of restricted solid-on-solid (RSOS) lattice models and their
fusion hierarchies [KP].
\medbreak

{\bf Corollary~2.7.}~~We have
$$\eqalignno{
&s(l,2,N)+s(N-1,2,N-2)-s(m-1,2,N-2)= &(2.14)\cr
&=c+1-24\Delta +6(l-|m|),~~m\in{\bf Z},~~
0\le l\le N, }
$$
where
$$c={2(N-1)\over N+2}~~{\rm and}~~\Delta ={l(l+2)\over 4(N+2)}-{m^2\over 4N}
\eqno (2.15)
$$
are the central charge and conformal weights of ${\bf Z}_N$ parafermion
theories  [FZ]. The members of (2.15) may be also realized as the central
charge and conformal weights of fusion $N+1$-state RSOS($p,p$) lattice models
[DJKMO], [BR] on the regime I/II critical line. Note that physical constraints
$$|m|\le l,~~m\equiv l~({\rm mod}~2)$$
for value of $m$ in (2.15) are equivalent to a condition that ``remainder
term'' in (2.14), namely, $6(l-|m|)$, must be belongs to $12{\bf Z}_+$.
\medbreak

{\bf Corollary~2.8.}~~Let us fix the positive integers $k,~~p=1,2,\ldots$
(the fusion level), $j_1$ and $j_2$ such that
$0\le j_1\le k,~~0\le j_2\le k+l$.

Let us put $r_0=p\left\{ \displaystyle{{j_1-j_2\over p}}\right\}$.
Then we have
$$\eqalignno{
&s(j_1,2,k)+s(r_0,2,p)-s(j_2,2,k+p)=& (2.16)\cr
&~~~ \cr
&=c-24\Delta +12~{(j_1-j_2)(p+j_1-j_2)
+r_0(p-r_0)\over 2p},}
$$
where
$$\eqalignno{
&c={3p\over p+2}\left( 1-{2(p+2)\over (k+2)(k+p+2)}\right), &(2.17)\cr
&~~~ \cr
&\Delta ={[(k+p+2)(j_1+1)-(k+2)(j_2+1)]^2-p^2\over 4p(k+2)(k+p+2)}+
{r_0(p-r_0)\over 2p(p+2)}}
$$
are the central charge and conformal weights of the fusion $(k+p+1)$-state
RSOS($p,p$) latice models [KP] on the regime III/IV critical line. It is easy
to see that ``remainder term'' in (2.16) belongs to $12{\bf Z}_+$. Note also
that the fusion RSOS($p,q$) lattice models, obtained by fusing~ $p\times q$~
blocks of face weights together, are related to coset conformal fields
theories obtained by the Goddard-Kent-Olive (GKO) construction [GKO]. Namely,
$c$ and $\Delta$ in (2.17) are the central charge and conformal dimensions
of conformal field theory, which corresponds to the coset pair [GKO]
$$\matrix{
&&A_1&\oplus& A_1&\supset& A_1\cr
{\rm levels}&&k&&p&&k+p\cr}
$$
%\vskip 0.8cm
\vfill\eject
%\bigbreak

{\bf \S 3.~~$A_1$-type~ dilogarithm~ identities.}
\medbreak

As is well-known [Le], the Rogers dilogarithm function $L(x)$ admits a
continuation on all complex plane ${\bf C}$. Follow [Le], [KR] we define
a function
$$\eqalignno{
L(x,\theta ):&=
-{1\over 2}\int_0^x{\log (1-2x\cos\theta+
x^2)\over x}dx+{1\over 4}\log |x|\cdot\log (1-2x\cos\theta +x^2)=\cr
&=ReL(xe{^{i\theta}}),~~x,\theta\in\bf R & (3.1) }
$$
Our proof of Theorem 2.1 is based on a study of properties of the function
$L(x,\theta )$.
\medbreak

{\bf Proposition~3.1.}~~ For all real $\varphi,~\theta$ we have
$$\eqalignno{
L\Big (\Big ({\sin\theta\over\sin\varphi}\Big )^2\Big )&=
%2\varphi\cdot\theta
\pi^2\biggl\{{\overline B}_2\left({\theta+\varphi\over\pi}\right) -
{\overline B}_2\left({\varphi\over\pi}
\right) -{\overline B}_2\left({\theta\over\pi}\right) +{1\over 6}\biggr\}+\cr
&+2L\Big (-{\sin (\varphi -\theta )\over\sin\theta},\varphi\Big )-
2L\Big (-{\sin\varphi\over\sin\theta},\varphi+\theta\Big ). & (3.2)\cr}
$$
Before proving a Proposition 3.1 let us give the others useful properties
of function (3.1) (compare with [Le]).
\medbreak

{\bf Lemma~3.2.}
$$\eqalignno{
(i)~~~~&L(x,0)=L(x),~~L(-x,\varphi )=L(-x,\pi -\varphi ) & (3.3) \cr
(ii)~~~&L(x,\varphi )=L(x,2\pi k\pm\varphi ),~~k\in {\bf Z } & (3.4) \cr
(iii)~~&L(-1,\varphi)=\pi^2{\overline B}_2\left({\varphi\over 2\pi}+
{1\over 2}\right), \cr
&L(1,\varphi)=\pi^2{\overline B}_2\left({\varphi\over 2\pi}\right) & (3.5) \cr
(iv)~~~&L(x,\varphi )+L(x^{-1},\varphi )=2\pi^2{\overline B}_2\left({\varphi
\over
2\pi}\right),
{}~~x>0 \cr
&L(-x,\varphi )+L(-x^{-1},\varphi )=
2\pi^2{\overline B}_2\left({\varphi\over 2\pi}+{1\over 2}\right),
{}~~x<0~~~~~~~~~~~~~~~~~~~~~~~~& (3.6) \cr
(v)~~~~&L(0,\varphi)=0,~~L(+\infty ,\varphi )=
2\pi^2{\overline B}_2\left({\varphi\over 2\pi}
\right), \cr
&L(-\infty ,\varphi )=2\pi^2{\overline B}_2\left({\varphi +\pi\over
2\pi}\right)
& (3.7) \cr
(vi)~~~&L(2\cos\varphi ,\varphi )=\pi^2\biggl\{ {\overline B}_2\left({\varphi
\over \pi}\right) +{1\over 12}\biggr\} & (3.8) \cr
(vii)~~&L(x^n,n\varphi)=n\sum_{k=0}^{n-1}L\biggl(x,~\varphi +
{2k\pi\over n}\biggr),~~
x\in{\bf R}_+ ,\cr
&L(x^n)=n\sum_{k=0}^{n-1}L\left( x\cdot \exp{2k\pi i\over n}\right)
,~~x\in (0,1).
& (3.9) \cr}
$$
More generally (Rogers' identity [Ro])
$$L(1-y^n)=\sum_{k=1}^n\sum_{l=1}^n\biggl[ L(\lambda_k/\lambda_l)-
L(x_k\lambda_l)\biggr] ,$$
where $\{ x_k\}^n_{k=1}$ are the roots of the equation
$$ 1-y^n=\prod^n_{k=1}(1-\lambda_kx).$$

Proof of the Theorem 2.1 for the case $n=2$.~~
If we substitute $\theta=m\varphi $ in (3.2) then obtain
$$\eqalignno{
&L\Bigg (\left({\sin m\varphi\over\sin\varphi}\right)^2\Bigg )=
\pi^2\biggl\{ {\overline B}_2\biggl({(m+1)\varphi\over\pi}\biggr)-
{\overline B}_2\biggl({m \varphi\over \pi}\biggr)-
{\overline B}_2\biggl({\varphi\over \pi}\biggr)+{1\over 6}\biggr\} +\cr
&+2L\biggl(-{\sin ((m-1)\varphi )\over\sin\varphi },m\varphi\biggr)-
2L\biggl(-{\sin m\varphi \over \sin\varphi},(m+1)\varphi \biggr) & (3.10)}
$$
Futher let us introduce notations
$$ f_m(\varphi ):=1-{Q_{m-1}(\varphi )Q_{m+1}(\varphi )\over Q^2_m(\varphi )}
={1\over Q_m^2(\varphi )}.$$
Then using (3.10) we find
$$\eqalignno{
&\sum^r_{m=1}L(f_m(\varphi ))=
-2\biggl\{ L\biggl( -Q_r(\varphi ),(r+2)\varphi \biggr)
-{\pi^2\over 6}\biggr\}+ &(3.11)\cr
&+\pi^2\left\{ {\overline B}_2\left({(r+2)\varphi\over\pi}\right)
-{1\over 6}\right\} +(r+2)\pi^2\left\{ {1\over 6} -
{\overline B}_2\left({\varphi\over\pi}\right)\right\} -
{\pi^2\over 2}.}
$$
Now let us put $\varphi =\displaystyle{{(j+1)\pi\over r+2}},
{}~~0\le j\le r+1$. Then
$Q_r(\varphi )=(-1)^j$ and it is clear from (3.3) and (3.4) that
$$ L\biggl( (-1)^{j+1},~(j+1)\pi \biggr) =L(1)={\pi^2\over 6}.$$
\qed

Note that polynomials $Q_m:=Q_m(\varphi )$ satisfy the following recurrence
relation
$$Q_m^2=Q_{m-1}Q_{m+1}+1,~~Q_0\equiv 1,~~m\ge 1,$$
where as the polynomials $y_m:=y_m(\varphi )=
Q_{m-1}(\varphi )\cdot Q_{m+1}(\varphi )$ satisfy the following one
$$ y_m^2=(1+y_{m-1})(1+y_{m+1}),~~y_0\equiv 0,~~m\ge 1.$$

Now we propose a generalisation of (3.12). Given a rational number
$p$ and decomposition of $p$ into the continued fraction
$$p=[b_r,b_{r-1},\ldots ,b_1,b_0]=b_r+{1\over\displaystyle b_{r-1}+
{\strut 1\over\displaystyle\cdots +{\strut 1\over\displaystyle b_1+
{\strut 1\over
\displaystyle b_0}}}}.\eqno (3.12)
$$
We will assume that $b_i>0$ if $0\le i<r$ and $b_r\in{\bf Z}$.
Using the decomposition (3.12) we define the set of integers $y_i$ and $m_i$:
$$\eqalignno{
&y_{-1}=0,~~y_0=1,~~y_1=b_0,\ldots ,y_{i+1}=y_{i-1}+b_iy_i,~~0\le i\le r,
&(3.13)\cr
&m_0=0,~~m_1=b_0,~~m_{i+1}=|b_i|+m_i,~~0\le i\le r.}
$$

The following sequences of integers were first introduced by Takahashi and
Suzuki [TS]:~~ $r(j)=i$, if $m_i\le j <m_{i+1},~~0\le i\le r$,
$$n_j=y_{i-1}+(j-m_i)y_i,~~{\rm if}~~m_i\le j<m_{i+1}+\delta_{i,r},~~0\le
i\le r.
$$
Finally we define a dilogarithm sum of ``fractional level $p$'':
$$
\sum_{j=1}^{m_{r+1}}(-1)^{r(j)}L\Bigg (\left({\sin y_{r{(j)}}\theta\over
\sin (n_j+y_{r{(j)}})\theta}\right)^2\Bigg ):=
(-1)^r{\pi^2\over 6}s(k,2,p),\eqno (3.14)
$$
where $\theta =\displaystyle{{(k+1)\pi\over y_{r+1}+2y_r}}$
\medbreak

{\bf Proposition~3.3.}~~We have
$$s(k,2,p)={3p\over p+2}-6{k(k+2)\over p+2}+6{\bf Z}. \eqno (3.15)
$$
%\medbreak

{\bf Conjecture 3.4.}~~For all positive $p\in{\bf Q}$ the remainder term in
(3.15) lies in $6{\bf Z_+}$.
\medbreak

{\bf Corollary 3.5.}~~Let us fix the positive integers $l=1,2,3,\ldots$ (the
fusion level), $p>q,~~j_1$ and $j_2$. Then
$$s\left( j_1,2,{ql\over p-q}-2\right) +s(r_0,2,l)-s\left( j_2,2,{pl
\over p-q}-2\right) =c-24\Delta +6{\bf Z},\eqno (3.16)
$$
where $r_0=l\cdot\left\{\displaystyle{{j_1-j_2\over l}}\right\}$ and
$$\eqalignno{
&c={3l\over l+2}\left( 1-{2(l+2)(p-q)^2\over l^2pq}\right),&(3.17)\cr
&\Delta ={[p(j_1+1)-q(j_2+1)]^2-(p-q)^2\over 4lpq}+{r_0(l-r_0)\over
2l(l+2)}}
$$
are the central charge and conformal dimensions of RCFT, which corresponds
to the coset pair [GKO]
$$\matrix{
&&A_1&\oplus& A_1&\supset& A_1\cr
& & & \cr
{\rm levels}&&{\displaystyle ql\over\displaystyle p-q}-2&&l
&&{\displaystyle pl\over\displaystyle p-q}-2\cr}
$$
\vskip 0.5cm
%\medbreak

{\bf Acknowledgements.} Most of this work was done during a visit to the
Isaac Newton Institute for Mathematical Sciences.
I thank W.Nahm, E.K.Sklyanin,\break
F.Ravanini and  E.Corrigan for very helpfull discussions.
%\vskip0.4cm
\vfill\eject

{\bf Riferences.}

\medbreak

\item{[Bi]} A.Bilal, Nucl.Phys., B330, (1990), 399.
\item{[BPZ]} A.Belavin, A.Polyakov, A.Zamolodchikov, J.Stat.Phys., 34,
(1984), 763; Nucl. Phys., B241, (1984), 333.
\item{[BR]} V.V.Bazhanov, N.Yu.Reshetikhin, Prog.Theor.Phys., 102, (1990),
suppl.
\item{[CR]} P.Christe, F.Ravanini, $G_N\otimes G_L/G_{N+L}$ conformal field
theories and their modular invariant partition functions, Int.J.Mod.Phys.,
A4, (1989), 897.
\item{[DR]} P.Dorey, F.Ravanini, Staircase model from affine Toda field
theory, prepr.\break SPhT/92-065.
\item{[FL]} V.A.Fateev, S.Lukyanov, Int.J.Mod.Phys.,A3, (1988), 507.
\item{[FQS]} D.Friedan, Z.Qiu, S.Shenker, Phys.Rev.Lett., 52, (1984), 1575.
\item{[GKO]} P.Goddard, A.Kent, D.Olive, Phys.Lett., B152, (1985), 105.
\item{[GM]} I.M.Gelfand, R.D.MacPherson, Advan. in Math., 44, (1982), 279.
\item{[Ka]} V.G.Kac, Infinite dimensional Lie algebras, Cambridge Univ.Press.
1990.
\item{[Ki1]} A.N.Kirillov, Zap.Nauch.Semin. LOMI, 164, (1987), 121.
\item{[Ki2]} A.N.Kirillov, Dilogarithm identities, partitions and spectra in
conformal field theory, in preparation.
\item{[KN1]} A.Kuniba, T.Nakanishi, Level-rank duality in fusion RSOS models,
prepr. 1989.
\item{[KN2]} A.Kuniba, T.Nakanishi, Spectra in conformal field theories from
Rodgers dilogarithm, prepr.SMS-042-92, 1992.
\item{[KP]} A.Klumper, P.Pearce, Physica, A183, (1992), 304.
\item{[KR]} A.N.Kirillov, N.Yu.Reshetikhin, Zap.Nauch.Semin. LOMI, 160,
(1987), 211.
\item{[Ku]} A.Kuniba, Thermodynamics of the $U_q(X_r^{(1)})$ Bethe ansatz
system with $q$ a root of Unity, ANU, prepr. SMS-098-91.
\item{[KW]} V.Kac, M.Wakimoto, Adv.Math., 70, (1988), 156.
\item{[Le]} L.Lewin, Polylogarithms and associated functions, (North-Holland,
1981).
\item{[MSW]} P.Mathieu. D.Senechal, M.Walton, Field identification in
nonunitary diagonal cosets, prepr. LETH-PHY-9/91.
\item{[Na]} W.Nahm, Dilogarithm and W-algebras, Talk given at the Isaac
Newton Inst., Cambridge, Sept. 1992.
\item{[NRT]} W.Nahm, A.Recknagel, M.Terhoeven, Dilogarithm identities in
conformal field theory, BONN-prepr., hepth 9211034.
\item{[Ro]} L.J.Rogers, Proc.London Math Soc., 4, (1907), 169.
\item{[RS] } B.Richmond, G.Szekeres, J.Austral.Math.Soc., A31, (1981), 362.
\item{[SA]} H.Saleur, D.Altshuler, Nucl.Phys., B354, (1991), 579.
\item{[TS]} M.Takahashi, M.Suzuki, Prog.Theor.Phys., 48, (1972), 2187.
\item{[Z]} Al.B.Zamolodchikov, Nucl.Phys., B342, (1990), 695.
\item{[Za]} D.Zagier, The remarkable dilogarithm (Number theory and related
topics. Papers presented at the Ramanujan colloquium, Bombay, 1988, TIFR).

%
%-------------------------------------------------------------------
%
\end